# Towards a conceptual framework of direct and indirect environmental effects of co-working

Bhavana Vaddadi

Integrated Transport Research Lab
KTH Royal Institute of Technology
Stockholm, Sweden
bhavana@kth.se

Jan Bieser

Department of Informatics
University of Zurich
Zurich, Switzerland
jan.bieser@ifi.uzh.ch

Johanna Pohl

Center for Technology and Society
Technische Universität Berlin
Berlin, Germany
pohl@ztg.tu-berlin.de

Anna Kramers

Department of Strategic Sustainability Studies
KTH Royal Institute of Technology
Stockholm, Sweden
kramers@kth.se

Teo Enlund

Product and Service Design unit, Department of Machine
Design,
KTH Royal Institute of Technology
Stockholm, Sweden
teoe@kth.se

*Abstract— Through virtual presence, information and communication technology (ICT) allows employees to work from places other than their employer's office and reduce commuting-related environmental effects (telecommuting). Co-working, as a form of telecommuting, has the potential to significantly reduce commuting and is not associated with deficits of working from home (e.g. isolation, lack of focus). However, environmental burden might increase through co-working due to the infrastructure required to set-up and operate the co-working space and potential rebound effects. In this paper, we (1) develop a framework of direct and indirect environmental effects of co-working based on a well-known conceptual framework of environmental effects of ICT and, (2) apply the framework to investigate the case of a co-working living lab established in Stockholm. Based on actual data of the co-working space and interviews conducted with participants, we roughly estimate associated energy impacts. Results show that energy requirements associated with operating the co-working space can counterbalance commute-related energy savings. Thus, in order to realize energy savings co-working should be accompanied with additional energy saving measures such as a net reduction of (heated) floor space (at the CW space, at the employer's office and the co-workers home) and use of energy-efficient transport modes.*

## I. INTRODUCTION

As cities continue to expand, people have started to move further away from the city centers due to housing shortages and ever-increasing rents making commuting a physical and a mental burden. Due to an unreliable transportation system and heavy dependence on personal vehicles, millions of people spend long hours commuting to and from work [1].

In 2011, roughly 38% of commuters in Stockholm were using private vehicles to commute to and from work while 25% used public transport [2]. In addition, car ownership and vehicular travel is ever increasing [3]. Besides its environmental impacts, commuting causes congestion during peak hours and has significant effects on individuals' well-being [4]. Hence, there is a dire need to adopt sustainable travel modes. Information and communication technology (ICT) has transformed our existing patterns of production and consumption with consequences for the environment [5], [6], [7], [8] . Telecommuting, working remotely and collaborating with colleagues and partners by means of ICT, has the potential to reduce commuting-related environmental impacts. While a lot of research has been conducted on environmental impacts of remote work from home, little is known about the environmental impacts of working from a telecommuting center.

A specific case of telecommuting centers are co-working (CW) spaces. CW "describes any situation where two or more people are working in the same place together, but not for the same company"[9, p. 3]. CW spaces are "shared workplaces utilized by different sorts of knowledge professionals […] working in various degrees of specialization in the vast domain of the knowledge industry" [10, p. 194]. CW holds the potential to significantly reduce environmental impacts associated with commuting and is not associated with deficits of working from home (e.g. isolation, lack of focus). In order to realize these benefits, the choice of location of the CW space is in particular critical [11].

However, CW can also increase environmental burdens, for example through required infrastructure to set-up and operate the CW. It can also lead to rebound effects, if employees spend time and money saved on commuting on other activities, goods and services that are associated with environmental impacts [12]. In order to draw more specific conclusions about whether CW can contribute to an overall reduction in resource consumption, and which factors are particularly relevant, a more precise analysis is necessary [11], [13], [14], [15].

One approach that has gained momentum in sustainability research is to test potentially sustainable innovations in living labs [16]. In living labs, technically and behaviorally data can be collected in a real-life setting and later be used for environmental assessment [17]. Within Mistra SAMS[1], a research project on sustainable transport in Sweden, a living lab CW space has been set up in the south of Stockholm (Tullinge) and is in operation since January 2019. As of February 2020, out of 60 recruited participants, about 44 employees who live close to the CW space regularly work from there and can potentially avoid lengthy commutes to their employers' offices.

In this paper, we (1) develop a conceptual framework of the diverse environmental impacts of CW, and (2) apply the framework to investigate environmental impacts associated with the CW living lab in Stockholm. Thereby, we provide a systematic overview of potential positive and negative environmental impacts of CW. We hope this can provide first insights on environmental impacts of CW and stimulate further research on CW and other promising ICT applications, which is required to harness the potential to avoid environmental burdens and mitigate negative impacts of increasing ICT use.

The paper is organized as follows: Materials and methods are described in Section 2. The conceptual framework of environmental effects of CW is presented in Section 3, followed by the application of the framework to the CW case in Stockholm in Section 4. We end with a discussion and conclusion and identify potential for future research in Section 5.

## II. MATERIALS AND METHODS

To develop a conceptual framework reflecting the environmental effects of CW, we use the framework of environmental effects of ICT by Hilty and Aebischer [8] and adapt it to the specific case of CW. The well-known and frequently applied taxonomy of environmental effects of ICT was introduced by Berkhout and Hertin [6] at first and has been revised several times since then [8], [14], [15]. The framework distinguishes three layers of environmental effects of ICT:

1. Direct environmental effects through production, use and disposal of ICT
2. Enabling effects of ICT use through the application of ICT in other sectors (the effects result from changes in production and consumption processes)
3. Structural impacts through ICT-induced changes of existing structures and institutions

This framework is useful to investigate the specific case of CW because,…

[1] Mistra Sustainable Accessibility and Mobility Services (Mistra SAMS) is a research program hosted and managed by KTH Royal Institute of Technology, Sweden in close cooperation with the Swedish National Road and Transport Research Institute (VTI) funded by Mistra, The Swedish Foundation for Strategic Environmental Research.

- CW is a specific use case of ICT as explained in the introduction,
- CW requires production, operation and disposal of infrastructures (e.g. CW space, ICT equipment), processes which cause environmental impacts (layer 1),
- CW can change existing production and consumption patterns (e.g. avoiding work-related travel or changing collaboration methods among colleagues - layer 2), and,
- CW can fundamentally affect the nature and location of work as well as transport habits at a societal level if it is adopted at a larger scale (e.g. through diminishing of central business districts - layer 3).

To adapt the framework, we applied the universally defined environmental effects of ICT to the specific case of CW. In a second step, we apply the framework to roughly estimate environmental impacts associated with the CW living lab in Stockholm. Wherever possible we use actual data collected in the CW living lab. We (1) collected technical data of the CW space, such as floor space and equipment used, (2) interviewed participants on their everyday life, travel and work patterns, and, (3) collected daily time-use data (time spent on 'travel', 'work', 'everyday chores' and 'leisure' time; use of transport modes) for three succeeding weeks by asking participants to fill out time-use diaries. Data collection took place from September until November 2019. As the living lab is still in operation and data collection is still ongoing, we cannot estimate some effects and in some cases have to use publicly available statistics about Stockholm or make reasonable assumptions.

## III. CONCEPTUAL FRAMEWORK OF ENVIRONMENTAL EFFECTS OF CO-WORKING

The framework, which describes direct, indirect and systemic environmental effects of CW, is shown in *Fig. 1*. The first layer, "Technology: Required infrastructure", describes the environmental effects of building, operating and maintaining infrastructures required for CW (e.g. CW space, video conferencing systems, parking places, etc.). The second layer, "Application: Working at the CW space", describes the environmental effects due to individual co-workers or organizations adopting to working at the CW space instead of the employer's office or from home. This directly affects the use of office space, transport infrastructure, and ICT equipment. In addition, behavioral changes, due to changing work and travel practices are possible. For example, employees might spend money and time saved on commuting on other activities that are associated with their own environmental impacts (patterns known as income and time-use rebound effects).

The third layer, "System: structural change", describes the environmental effects of a system transformation towards CW. It leaves the level of individual co-workers or organizations and focuses on environmental consequences of a transformation towards society-wide CW culture. Such effects include changes to working cultures, ways of communication, lifestyles or land use patterns, which only occur if a critical mass of society switches from conventional working habits to CW.

In the following, we describe each layer in some detail. In the framework, we included effects described in literature and observed during operation of the CW living lab. Still, effects beyond the ones we describe can exist.

*A. Technology layer*

The direct environmental effects of building, operating and maintaining CW spaces are by definition unfavorable environmental effects as they all require resources and cause emissions, but do not avoid anything yet. Main environmental impacts associated with building and operating a CW space are caused by facilities (main offices, auxiliary rooms, parking) and equipment (ICT equipment and infrastructure, office furniture) (TABLE I).

Environmental impacts caused throughout the life cycle of facilities and equipment are caused by the construction of facilities and production of equipment (production phase), the operation of these (use phase) and processes at their end-of-life (EoL phase). As for the production phase, the construction of CW spaces and production of ICT equipment, furniture and other required equipment cause environmental impacts.

With regard to facilities, energy consumption during the operational phase is of great relevance [18] and can be divided into energy for heating, cooling and lighting. Use phase energy demand in office buildings can be estimated proportional to office space [19]. With increasing adoption of energy-efficient building technologies (e.g. improved insulations) the relative importance of the construction phase increases.

With regard to ICT equipment, the relevance of the production phase depends on the type of the equipment, the service life and energy efficiency of the devices. The smaller and more efficient the devices, the more important is the use phase [20] .

With regard to ICT infrastructure, communication infrastructure (e.g. networks) as well as servers (or data centers) are most relevant. Overall, the total number of devices used is decisive for the total environmental impacts. The main target on this layer is to reduce the relative effects per co-worker that stem from constructing, operating, and maintaining facilities and equipment. Amongst others, this means to minimize required office space and to aim for high occupancy rates.

*B. Application layer*

The environmental effects resulting from running and using the CW space can work in both directions (e.g. reducing and increasing resource use). Main environmental impacts of CW are caused by changes to the process/use of (1) travel, (2) office space, and, (3) ICT equipment and infrastructure. The main driver of environmental impacts on this layer is changes to the floor space at the employer's office and the reduction of commuting distance.

As discussed in the introduction, CW spaces that are close to the employees' homes can contribute to a reduction in commuting time and commuting distance. This is the case, if trips to the CW space replace commuting trips to work. If working from the CW space replaces working from home, commuting time and commuting distance increases instead.

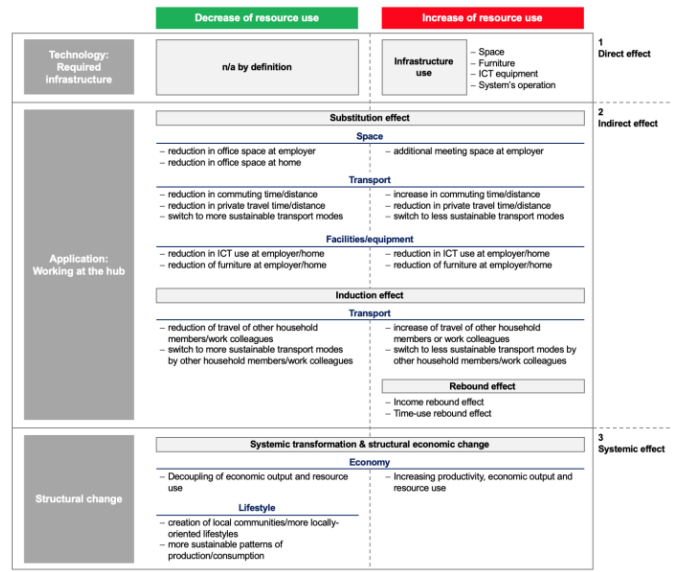

Fig. 1. Framework of environmental effects of co-working (based on [8] and [15]) "[Question to conference organizers: Where should we place to figure in order to achieve a larger font size?]"

TABLE I FACILITIES AND EQUIPMENT IN THE CO-WORKING SPACE.

| Facilities | Main use area | • Workplaces<br>• Meeting rooms<br>• Telephone rooms<br>• Event Space |
|---|---|---|
| | Auxiliary areas | • Kitchen<br>• Bathrooms<br>• Parking space |
| Equipment | ICT equipment and infrastructure | • End-user devices (screens, printers, white boards)<br>• Infrastructure (e.g. network, servers)<br>• Conferencing equipment |
| | Office furniture | • Desks<br>• Chairs |
| | Other | • Coffee machine<br>• Cleaning equipment<br>• … |

If, before the adoption of CW, private activities such as library visits, meeting friends or shopping had been combined with commute trips, CW can also induce additional trips. Further, changes in commuting can lead to a change in transport modes used (modal split). For example, for shorter commutes people might consider taking the bike instead of the car. However, people might also increase their use of cars for shorter commute trips, because the opportunity cost of taking the car instead of public transport are less significant (in public transport people can do other activities).

Furthermore, the saved travel money can be used for other purposes (income rebound effects) and thus contribute to an increased use of resources. Finally, co-workers can spend saved commuting time on other activities that are associated with environmental impacts (time-use rebound effects [21]). Work-ing from CW spaces has the potential for a reduction of office space at the employer's office and the employee's home (e.g. by implementing desk sharing at the employer's office). However, if these office spaces are not sufficiently reduced, CW can have a net increasing effect on office space. Also, CW might increase demand for meeting space at the employer's office, which is required to communicate with co-workers. Employers adopting CW might also require additional ICT equipment (e.g. for video conferencing). Use of (additional) ICT equipment should also be minimized in order to reduce associated environmental impacts.

The main target on this layer is to promote desired and mitigate undesired effects. The effect of CW on (heated) floor space (at the employer and at the co-worker's home), the average change in commuting distance of co-workers, thus, the location of the CW space (central, sub-urban, close to the co-workers houses), and the transport modes used, seem to be the most important drivers of the environmental impacts on the application layer.

### C. Structural change layer

Structural effects of CW are effects that occur if CW is adopted at a larger scale. For example, given that CW reduces time spent on commuting and adds flexibility to time and place of work, it may influence families' decisions regarding where to live, jobs, and investments in their dwellings [22], [23], [24]. In the long-term this can also change land-use patterns, e.g. towards "more decentralized and lower-density land use patterns" [25, p. 12]. CW from local CW spaces at a larger scale can also change the nature of work and would reduce demand for major office buildings in business districts, which then could be used for other purposes.

Finally, CW can also change traffic streams and demand for transport in general. Rebound effects occur also on the structural layer. For example, if CW increases the productivity of an industry and stimulates growth; this can lead to increase in resource consumption and emissions (economy-wide rebound effect) [13], [26]. Structural effects of CW depend on many variables in the broader societal and economic system and are therefore difficult to predict. A long-term CW strategy at a larger scale needs to identify potential structural effects and promote CW schemes that foster environmentally favorable structural effects and mitigate unfavorable ones.

## IV. CASE STUDY: ENVIRONMENTAL EFFECTS OF A CO-WORKING SPACE IN STOCKHOLM

### A. Introduction to the co-working space in Stockholm

Situated in Tullinge, a suburb in the south of Stockholm, the CW space is an experimental living lab set up to observe a wide range of effects of having a workplace close to the home of the participants. The CW space integrates various accessibility and mobility services to participants to allow them to book, plan, and travel. It offers an activity-based workplace close to co-workers' homes (requirement for participating was to live close to the CW space), gives access to 3 electric bikes (2 electric bicycles and 1 electric cargo bicycle) for free and a peer-to-peer carpooling scheme.

It is equipped with 14 workplaces, which can be booked via an online application, a well-equipped conference room for eight people, as well as three rooms for telephone or video calls. This experimental CW space acts as a platform to bring together a range of actors such as citizens, researchers, business and public authorities to create, validate, and test new mobility and accessibility technology and services in a real-life context. The CW space is in operation since January 2019 and as of February 2020 44 out of 60 participants regularly work there.

### B. Co-working impacts on time use, transport and energy requirements

We used the results of the time-use diaries of 20 co-workers who work for an IT company in Kista, north of Stockholm to estimate commute-related energy savings. Because living close to the CW space in the south of Stockholm was a requirement for participating, these co-workers significantly reduce their commuting time and distance on CW days compared to employer office days. We compare time spent on 'travel', 'work', 'everyday chores' and 'leisure' on days, when people work from the employer's office, from the CW space or from home (Fig. 2).

We also compared the (share of) time people spent in different transport modes on these days (

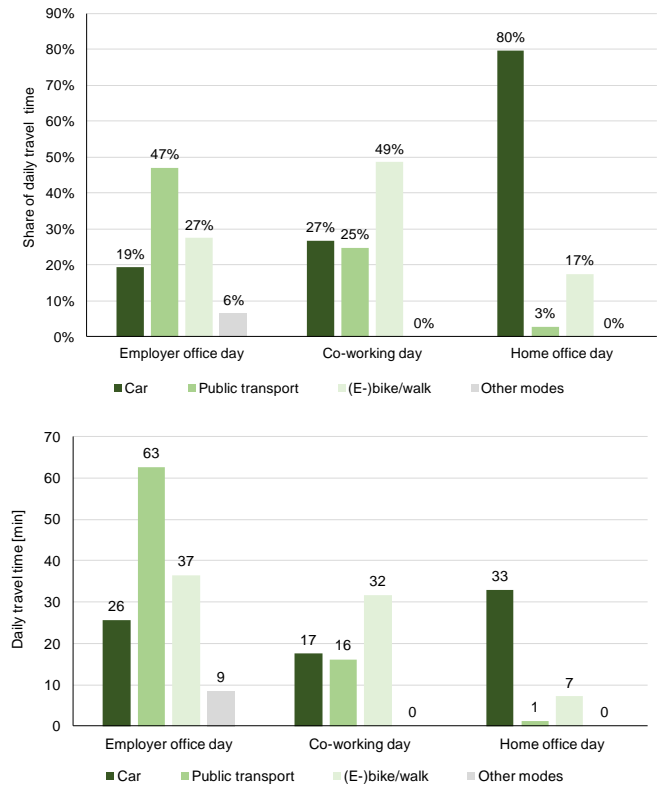

Fig. 3). We did not consider days, when people worked from other locations or from several locations on one day. We also excluded low quality data entries and untypical work days (work time lower than 4h; total recorded time lower than 8h; time difference between the recorded time spent on traveling and recorded time in specific transport modes is higher than

100min; these were two separate questions). This results in time-use data from 250 workdays.

### 1) Time spent on activities

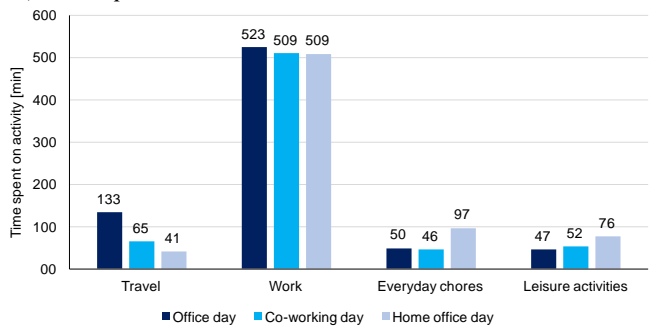

Fig. 2. Time spent on an activity by work location on that day.

'Travel' time is highest, when people work from the employer's office (133min) and decreases by 68min on CW days and 92min on home office days. Working time is also highest on days, when people work from the employer's office (522m) and decreases slightly on CW (-14min) and home office days (-14min). Time spent on 'everyday chores' and 'leisure' is highest on home office days and lower on days when people work from the employer's office or the CW space. Impacts on other activities (e.g. sleep) are also possible, but were not collected in the time-use diaries.

### 2) Used transport modes (modal split)

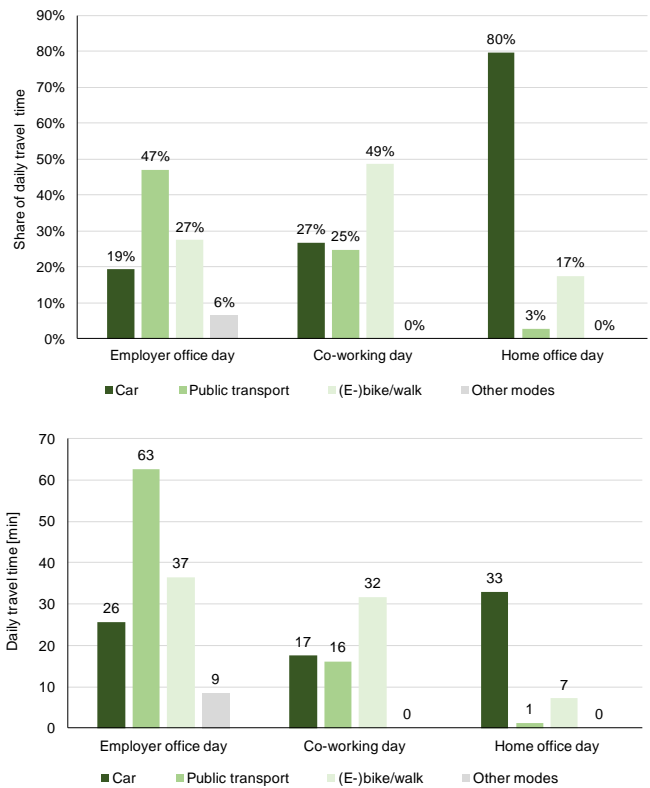

Fig. 3. Share of travel time (top) and absolute time (bottom) spent in different transport modes by work location on that day (other modes are for example boats).

On office days, people spent most time in public transport, followed by (e-)bike and walking (27%). Lowest amount of travel time is spent in the car (19%). In the interviews, we also asked participants about their commuting travel modes specifically. These results indicate that public transport is the preferred commute transport mode, followed by car transport. This indicates that biking and walking is rather done for private purposes. In contrast, on CW days, the share of travel time in the car and in public transport are almost equal, share of time spent biking and walking is very high. This indicates that people, who use public transport to commute to the employer's office, switch to biking, walking or commuting by car on CW days.

On home office days, the car is by far the preferred travel mode (80% of time spent in transport). On these days, people do not have to commute; thus, they use the car for private purposes. As absolute transport time is significantly lower on CW and home office days than on employer office days, absolute time spent in all transport modes is lower on CW days and home office days than on employer office days; except for car travel time, which is highest on home office days.

### 3) Estimation of energy impacts

Applying the framework of environmental effects of CW, we estimate the direct and indirect (no systemic) energy impacts of the CW living lab in Tullinge. To do so, we compare energy requirements on employer office days and CW days as well as on home office days and CW days. In our estimation, we include energy requirements associated with heating, cooling and lighting of the CW space' (direct effect), 'ICT equipment operated in the CW space' (direct effect) and 'changes in travel time' (substitution effect).

Due to lack of data, we do not estimate energy impacts due to changes in ICT, space and furniture use at the employer's office or at the employee's home, neither effects on behaviour of other household members or work colleagues (e.g. changes in travel). To some extent, changes in travel time include income and time-use rebound effect, as people can spend saved commuting cost and time on travel for other purposes.

All calculations are performed for one CW day of one co-worker in comparison to working at the employer's office or at home. The energy impacts are energy requirements associated with the operation of the CW space and fuel consumption for transport (use phase). Energy impacts associated with production of goods and services (e.g. production of car, construction of office building, and production of ICT equipment) are out of scope.

We use actual data from the CW living lab (time-use diaries; infrastructure data) and estimate energy consumption associated with it. TABLE II provides data on floor area, ICT equipment and the number of people in the CW space.

TABLE II CW FLOOR AREA, AMOUNT OF ICT EQUIPMENT USED IN THE CW SPACE AND NUMBER OF CO-WORKERS.

| Building | | |
|---|---|---|
| **ICT equipment** | Floor area co-working space [m$^2$] | 170 |
| | Workplaces | 14 |
| | Screens | 18 |
| | Desktop computers | 1 |
| | Printers | 1 |
| | TV sets | 1 |
| **Co-workers** | Total | 60 |
| | From IT-company in Kista (diaries avail.) | 20 |

To estimate energy impacts of heating, cooling and lighting of office space we used the floor space of the CW space and yearly energy requirements of standard office buildings according to the "Institut Wohnen und Umwelt" [27]. We divided energy impacts of heating, cooling and lighting of office space by the number of people working in the CW space and the number of workdays per year to estimate impacts per co-worker and CW day. Thereby, we assume that co-workers who work for other companies have the same CW patterns (number of CW days) as the co-workers working for the IT company in Kista.

For operation of ICT equipment, we used the number of devices in operation in the CW space and daily device energy requirements according to ecoinvent [28]. To estimate impacts per co-worker and CW day, we divided ICT equipment energy consumption by the number of workplaces at the CW space. We did not include network devices and one videoconferencing system due to lack of data.

To estimate energy impacts of changes in travel time, we used the results of the time-use diaries and interviews for travel time and modal split, direct energy requirements of fuel consumption and provisioning of travel modes according to mobitool [29] and average speed of transport modes [30].

Fig. 4 shows the estimated change in energy consumption due to one person working from the CW space for one day, compared to working from the employer's office or home. It shows that the main increase in energy requirements is due to heating, cooling and lighting (mainly heating and lighting, only few cooling) of CW office space (23.97 MJ) and to a small extent due to ICT equipment (2.03 MJ).

Compared to employer office days, reduction in travel leads to a reduction of travel-related energy impacts of 21.95 MJ; thus, energy impacts of reduction in travel and increase in heating, cooling and lighting of office space roughly cancel each other out. Compared to home office days, co-workers spend more time travelling on CW days; still travel-related energy consumption decreases. This is because on home office days, people rather use the car instead of other transport modes, whereas on CW days car and public transport have a similar, and, walking and biking, an even higher modal share. Still, travel-related energy reductions are higher if CW substitutes days at the employer's office and not at home.

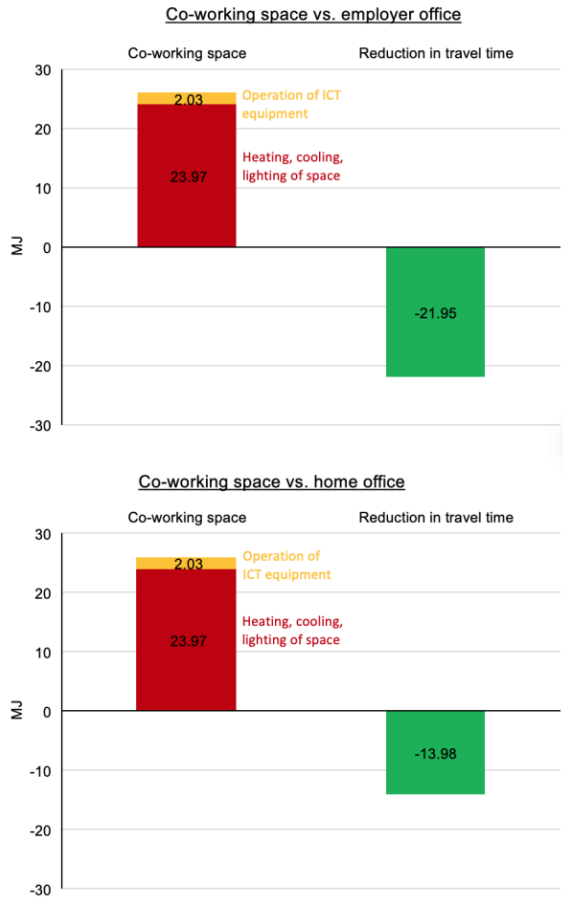

Fig. 4. Change in energy requirements on a CW workday compared to workday at the employers' office (top) or workday at home (bottom)

While energy requirements due to heating, cooling and lighting of the CW space are rather stable and do not fluctuate strongly with the utilization of the office space (e.g. changes in number of co-workers or CW days), the number of CW days is proportional to travel-related energy savings (e.g. one CW day avoids one commute, two CW days avoid two commutes,…). Thus, increasing the number of CW days (or the utilization of the CW space) is a good strategy to increase CW-related energy savings.

This estimation did not consider changes in energy consumption at home or at the employer's office. It is plausible to assume a decrease in these energy requirements, thus leading to a larger decreasing effect of CW. However, income and other rebound effects could compensate for the savings. We also could not consider interdependencies between weekdays and weekends. For example, people could systematically shift activities for which they require the car (e.g. shopping) from weekdays to weekends. This increases the car use on weekends, but total car use per week remains stable.

## V. DISCUSSION AND CONCLUSION

CW from a local CW space is a promising ICT use case to reduce transport demand and associated environmental impacts, while having a positive effect on well-being of employees (e.g. time savings). However, CW also causes environmen-tal impacts, for example through infrastructure required to operate CW spaces or through time-use rebound effects.

Based on an existing framework of environmental effects of ICT, we developed a conceptual framework of environmental effects of CW. The framework distinguishes environmental effects of CW on three layers: (1) direct effects through the infrastructure required to operate CW spaces, (2) indirect effects due to individual co-workers or organizations adopting CW (e.g. avoided commutes), and, (3) structural effects through a system transformation towards CW (e.g. fundamental changes in demand for transport and office space).

While direct effects are environmentally unfavourable by definition (they increase resource use), indirect effects and systemic effects can increase but also reduce resource use (e.g. by avoiding commute time or inducing additional travel for other purposes). Thus, net environmental effects depend on the magnitude of effects on all three layers and institutions should consider them when developing and adopting CW schemes.

In our case study of a CW living lab in Stockholm, we found that co-workers travelled most on employer office days, less on CW days and least when they worked from home. However, changes in travel mode can counterbalance this effect, as we found in our case study: On home office days, participants preferably travelled by car (energy-intensive travel mode); whereas on CW days people spent more time walking and biking and roughly the same amount in public transport as in cars.

In our case study, the energy required to operate the CW space and travel-related energy savings roughly counterbalance each other on employer office and CW days. Thus, CW does not lead to energy savings per se, but should be accompanied by additional energy savings measures, such as reduction of office space at the employer's office. In addition, increasing the number of CW days increases the number of avoided commutes and, thus, energy savings. To summarize, the main levers to realize energy savings through CW are a reduction of total travel time and distances (e.g. by choosing CW spaces close to home), use of sustainable transport modes, a net reduction of (heated) floor space (at the CW space, at the employer's office and the co-workers home) and a high utilization of the CW space. Our calculations have limitations and uncertainties regarding the extent of daily activities captured the energy intensity of activities and the consideration of structural effects. We focused on operational energy demand, thus environmental effects related to the production, construction and disposal of buildings, devices, vehicles and roads are not included in our estimation.

Furthermore, we presented our results in terms of energy associated with adapting CW. Environmental impacts beyond energy use (e.g. global warming potential or human toxicity) exist and need to be investigated to provide a full picture of environmental effects of CW. Future research should take a broader perspective both in terms of effects and activities included in the calculations and environmental impact categories and life cycle stages considered. If CW is adopted at a larger scale, systemic effects can lead to fundamental transformation of transport system and land use. These effects are difficult to estimate and further research is required. We encourage companies and researchers to experiment with CW and find ways to use CW for reducing environmental effects of transport, work and everyday life. The framework developed in this paper and the findings of the living lab can provide guidance for this.

ACKNOWLEDGMENT

This research is supported by the research program Sustainable Accessibility and Mobility Services – Mistra SAMS, funded by the Swedish Foundation for Strategic Environmental Research, Mistra, by the "Forschungskredit of the University of Zurich, grant no. FK-19-014" and by the German Federal Ministry of Education and Research within the project "Digitalization and Sustainability".